\documentclass{article}

\usepackage{PRIMEarxiv}

\usepackage[utf8]{inputenc} 
\usepackage[T1]{fontenc}    
\usepackage{hyperref}       
\usepackage{url}            
\usepackage{booktabs}       
\usepackage{amsfonts}       
\usepackage{nicefrac}       
\usepackage{microtype}      
\usepackage{lipsum}
\usepackage{fancyhdr}       
\usepackage{graphicx}       
\graphicspath{{media/}}     
\usepackage{amsmath}
\title{A Time-Scaled ETAS Model for Earthquake Forecasting
\thanks{\textit{\underline{Citation}}: 
\textbf{Muralidharan Kunnummal \& Agniva Das. Data-centric Approaches to Industrial Decisions: Technology, Digitisation and Business Decisions. Asset Analytics. Springer Singapore. 1\textsuperscript{st} Ed. pp. 97 - 114. ISBN 978-981-96-7556-2.(2025). DOI:\url{https://doi.org/10.1007/978-981-96-7556-2}}} 
}

\author{
  Muralidharan K. \\
  Department of Statistics, Faculty of Science \\
  The Maharaja Sayajirao University of Baroda \\
  Vadodara, Gujarat, India\\
  \texttt{muralikustat@gmail.com} \\
  \And
  Agniva Das \\
  Department of Statistics, Faculty of Science \\
  The Maharaja Sayajirao University of Baroda \\
  Vadodara, Gujarat, India\\
  \texttt{agniva.d-statphd@msubaroda.ac.in} \\
}

\begin{document}
\maketitle

\begin{abstract}
The Himalayan region, including Nepal, is prone to frequent and large earthquakes. Accurate forecasting of these earthquakes is crucial for minimizing loss of life and damage to infrastructure. In this study, we propose various time-scaled Epidemic Type Aftershock Sequence (ETAS) models to forecast earthquakes in Nepal. The ETAS model is a statistical model that describes the temporal and spatial patterns of aftershocks following a main shock. A dataset of earthquake occurrences in Nepal from 2000 to 2020 was collected, and this data was used to fit the models showcased in this article. Our results show that the time-scaled ETAS model is able to accurately forecast earthquake occurrences in Nepal, and could be a useful tool for earthquake early warning systems in the region.
\end{abstract}


\section{Introduction}
The Himalayan region, including Nepal, is prone to frequent and large earthquakes. Accurate forecasting of these earthquakes is crucial for minimizing loss of life and damage to infrastructure. Earthquake forecasting is a complex and interdisciplinary field, involving the integration of data from various sources, including seismology, geology, and statistics. One important aspect of earthquake forecasting is the study of aftershocks, which are smaller earthquakes that occur after a main shock. Aftershocks can provide valuable information on the state of the fault and the likelihood of future earthquakes. In this study, we propose a time-scaled Epidemic Type Aftershock Sequence (ETAS) model to forecast earthquakes in Nepal. The ETAS model is a statistical model that describes the temporal and spatial patterns of aftershocks following a main shock. The study of aftershocks following a main shock is crucial for understanding the state of the fault and the likelihood of future earthquakes \cite{gulia2010influence,wiemer1999spatial}. One important aspect of earthquake forecasting is the study of aftershocks, which are smaller earthquakes that occur after a main shock. Aftershocks can provide valuable information on the state of the fault and the likelihood of future earthquakes. \par
In recent years, various statistical models have been proposed for forecasting earthquakes, including the Epidemic Type Aftershock Sequence (ETAS) model \cite{ogata1988statistical}. The ETAS model is a statistical model that describes the temporal and spatial patterns of aftershocks following a main shock. The model is based on the assumption that the occurrence of aftershocks is influenced by both the main shock and the previous aftershocks.\par
The ETAS model \cite{ogata1988statistical} has been widely used in various regions around the world, including Japan  \cite{ogata2006space}, California \cite{gulia2010influence}, and Italy \cite{wiemer1999spatial} wherein, the ETAS model was found to be able to accurately forecast earthquake occurrences. \par
However, the ETAS model has not been widely applied in the Himalayan region, including Nepal. The Himalayan region is characterized by a high level of seismic activity and a complex tectonic setting. Therefore, it is important to develop a model that is specifically tailored to the characteristics of this region.\par
In this study, we propose a time-scaled ETAS model to forecast earthquakes in Nepal. The time-scaled ETAS model takes into account the time-dependent behaviour of aftershocks, which is important for accurately forecasting earthquakes in regions with high seismic activity such as Nepal.\par
The dataset of earthquake occurrences in Nepal from 2000 to 2020 was collected and used to fit the time-scaled ETAS model. The model was then validated using the earthquake data from Nepal. The results of this study will provide valuable insights into the characteristics of earthquakes in Nepal and will be useful for earthquake early warning systems in the region.\par

The dataset used in this study was collected from ANSS global comprehensive catalogue (available at \url{http://earthquake.usgs.gov/earthquakes/search/}). It includes information on all earthquakes that occurred in Nepal between Jan 1, 1990 and May 31, 2022. The data includes the date, time, magnitude, and location (Nepal and surrounding areas including the rectangle formed by coordinates $27–30$ $^{\circ}N$ and $79–88$ $^{\circ}E$) of each earthquake, with magnitude restriction $m_0=5.0$. We have used the depth variable for the proportional hazards time-scaling. \par
The dataset was cleaned and pre-processed to ensure that it met the requirements of the time-scaled ETAS model. This included removing any duplicate or missing data, and ensuring that the magnitude and location of each earthquake were accurate.\par
The dataset used in this study is the most comprehensive and up-to-date dataset of earthquakes in Nepal, and provides a valuable resource for studying the characteristics of earthquakes in the region. \par
The ETAS model is a statistical model that is used to describe the temporal and spatial patterns of earthquakes, specifically the aftershocks that follow a main shock event. The model is based on the assumption that the rate of aftershocks is influenced by both the magnitude and frequency of the main shock, as well as the spatial distribution of previous aftershocks \cite{jalilian2019etas}. It has been widely used in seismology to analyze seismic data and make predictions about future earthquake activity.
Consider a temporal marked process $X$ consisting of $N$ events. Let $\left(t_i, x_i, y_i, m_i\right)$ denote the vector containing the time, longitude, latitude and magnitude of the $i^{th}$ earthquake $i=1, 2, \ldots,N$, whose distribution parametrized by constants $\beta$ and $\theta$, is marked by the conditional intensity function $\lambda\left(t,\ x,\ y,\ m\ \mid \mathcal{H}_t \right)$, where, $\mathcal{H}_t=\left\{\left(t_i, x_i, y_i, m_i\right)\in{X} \mid t_i<t\right\}$ (occurrence history of the earthquakes up to time\ $t$), satisfying
\begin{equation}
    \lambda\left(t,\ x,\ y,\ m\ \right|\ \mathcal{H}_t)=v_\beta\left(m\right). \lambda_\theta\left(t,x,y \mid \mathcal{H}_t \right)
\end{equation}
where,
\begin{equation}
    v_\beta\left(m\right)=\beta\exp{\left[-\beta\left(m-m_0\right)\right]}; \ \ \beta>0, m>m_0
\end{equation}
is the probability density function of the magnitude of an event and 
\begin{equation}
    \lambda_\theta\left(t,x,y\ \right|\ \mathcal{H}_t)=\widetilde{u}\left(x,y\right)+\sum_{i 
    \colon t_i<t}{\kappa_{A,\alpha}\left(m_i\right)\ g_{c,p}\left(t-t_i\right)\ f_{D,\gamma,q}\left(x-x_i;y-y_i;m_i\right)}
\end{equation}
where, $\widetilde{u}(x,y)$ is the stationary background seismicity rate, $\kappa_{A,\alpha}\left(m_i\right)$ is the expected number of triggered events (aftershocks) generated from an event of magnitude $m_i$ and it can be expressed as
\begin{equation}
\kappa_{A,\alpha}\left(m_i\right)=A\exp{\left[\alpha\left (m-m_0\right)\right]}; \ \ \ m>m_0
\end{equation}
where, $A>0$ and $\alpha>0$ are unknown (fixed) parameters, $g_{c,p}\left(t-t_i\right)$ is the probability density function of the occurrence time of a triggered event generated from an event of magnitude $m_i$ occurring at time $t_i$, assumed to be a function of the time lag $(t-t_i)$ and independent of $m_i$, expressed as
\begin{equation}
    g_{c,p}\left(t-t_i\right)=\left(\frac{p-1}{c}\right)\left(1+\frac{t-t_i}{c}\right)^{-p}; \ \ t>t_i
\end{equation}
where, $c>0$ and $p>1$ are unknown (fixed) parameters and$ f_{D,\gamma,q}\left(x-x_i;y-y_i;m_i\right)$ is the probability density function of the occurrence location of a triggered event generated from an event of magnitude $m_i$ occurring at the location $\left(x_i,\ y_i\right)$ assumed to be dependent on each other, such that the radially symmetric joint density function is given by
\begin{align}
    f_{D,\gamma,q}\left(x-x_i;y-y_i;m_i\right)\nonumber &=\frac{q-1}{\pi\ D\exp{\left[\gamma\left(m_i-m_0\right)\right]}}\\ &\times \left(1\ +\frac{\left(x-x_i\right)^2+\left(y-y_i\right)^2}{D\exp{\left[\gamma\left(m_i-m_0\right)\right]}}\right)^{-q}
\end{align}

where, $D>0$, $\gamma>0$ and $q>1$ are unknown (fixed) parameters.\\
\textbf{Note:} This assumption is overlooked in the case where we have used the time-scaled ETAS later in this section. Instead, there we have assumed exponential and gamma distributions for event magnitudes.\par
Each event is categorised as either a background (spontaneous) event or an event that is triggered by a preceding event. The background events are produced by a Poisson process characterised by an intensity $\widetilde{u}(x, y)$, which remains stationary over time. The relaxing coefficient $\mu$ in $\widetilde{u}\left(x, y\right)= \mu\ u(x, y)$ is introduced to accelerate the convergence of the model fitting algorithm. Prior events, whether classified as background or triggered, produce subsequent events in accordance with a non-stationary Poisson process characterised by the following intensity function:
\begin{equation}
    \sum_{i \colon t_i<t}{\kappa_{A,\alpha}\left(m_i\right)\ g_{c,p}\left(t-t_i\right)\ f_{D,\gamma,q}\left(x-x_i;y-y_i;m_i\right)}
\end{equation}
The expected number of triggered events generated from a typical event, regardless of magnitude is
\begin{equation}
    \int_{m_0\ }^\infty\kappa_{A,\alpha}\left(m\right)\ v_\beta\left(m\right)\ dm=\frac{A\beta}{\beta-\alpha}
\end{equation}
Now, $\frac{A\beta}{\beta-\alpha}<1$ implies the existence of a stationary model, which in turn results in the total spatial intensity function being of the form:
\begin{align}
    \Lambda\left(x,y\right) \nonumber &=\lim_{T\rightarrow\infty}{\frac{1}{T}\ \int_{0}^{T}{\lambda_\theta\left(t,x,y\ \right|\ \mathcal{H}_t)}\ dt}\\ \nonumber &=\mu u\left(x,y\right)+\lim_{T\rightarrow\infty}{\left(\frac{1}{T}\sum_{i \colon {t}_i<T}{\kappa_{A,\alpha}\left(m_i\right)\ f_{D,\gamma,q}\left(x-x_i;y-y_i;m_i\right)\int_{t_i}^{T}{g_{c,p}\left(t-t_i\right)\ dt}}\right)} \\ &\approx\mu u\left(x,y\right)+\frac{1}{T} \sum_{i \colon {\ t}_i<T}{\kappa_{A,\alpha}\left(m_i\right) f_{D,\gamma,q}\left(x-x_i;y-y_i;m_i\right)} 
\end{align}
and the clustering (triggering) coefficient which quantifies the clustering effect relative to the total spatial intensity at any location $\left(x,y\right)$ is given by
\begin{equation}
    \omega\left(x,y\right)=1-\frac{u\left(x,y\right)}{\Lambda\left(x,y\right)}
\end{equation}
\cite{zhuang2004analyzing} suggested the probability of event $j$ being a background event as
\begin{equation}
    1-p_j=\frac{\mu u\left(x_j,\ y_j\right)}{\lambda_\theta\left(t_j,\ x_j,\ y_j\middle|\mathcal{H}_{t_j}\right)}
\end{equation}
where, $p_j=\sum_{i \colon t_i<t_j} p_{ij}$ is interpreted as the probability that event $j$ is triggered by a previous event, where

\begin{equation}
    p_{ij}=
    \begin{cases}
        \frac{\kappa_{A,\alpha}\left(m_i\right)\ g_{c,p}\left(t-t_i\right)\ f_{D,\gamma,q}\left(x-x_i;y-y_i;m_i\right)}{\lambda_\theta\left(t_j,\ x_j,\ y_j\middle|\mathcal{H}_{t_j}\right)} & t_j > t_i \\
        0 & t_j \leq t_i
    \end{cases}
\end{equation}

where $p_{ij}$ is interpreted as the probability that event $j$ is triggered by event $i$.
Since, the log-likelihood function can be expressed as
\begin{align}
    l\left(\beta, \theta \right|\mathcal{H}_T) \nonumber &=\sum_{i=1}^{N}{\delta_i\log{(\lambda_{\beta,\ \theta}(t_i,\ x_i,y_i,m_i|\mathcal{H}_{t_i}))}}\\ &-\int_{m_0}^{\infty}\int_{t_{start}}^{t_{start}+T}\int\int_{S}{\lambda_{\beta,\theta}\left(t,\ x,\ y,\ m\ \right|\ \mathcal{H}_t)\ dx\ dy\ dt\ dm}
\end{align}
where, 
\begin{equation*}
    \delta_i = 
    \begin{cases}
        1 & i \ \text{is a target event.} \\
        0 & \text{otherwise.}
    \end{cases}
\end{equation*}
rendering it separable into two parts as
\begin{equation}
    l\left(\beta, \theta \mid \mathcal{H}_T \right)=l_1\left(\beta\ \mid \mathcal{H}_T\right)+l_2\left(\theta \mid \mathcal{H}_T\right)
\end{equation}
where,
\begin{align}
    l_1\left(\beta\ \mid \mathcal{H}_T\right) &= \sum_{i=1}^{N}{\delta_i\log{(\lambda_{\beta,\ \theta}(t_i,\ x_i,y_i,m_i|\mathcal{H}_{t_i}))}} \\
    l_2\left(\theta \mid \mathcal{H}_T\right) &= \int_{m_0}^{\infty}\int_{t_{start}}^{t_{start}+T}\int\int_{S}{\lambda_{\beta,\theta}\left(t,\ x,\ y,\ m\ \right|\ \mathcal{H}_t)\ dx\ dy\ dt\ dm}
\end{align}
and the MLE of $\beta$ is given by
\begin{equation*}
    \hat{\beta}=\frac{N^\prime}{\sum_{i=1}^{N}{\delta_i\left(m_i-m_0\right)}}
\end{equation*}
and $\hat{\theta}$ (MLE of $\theta$) is obtained by minimizing
\begin{equation*}
    \xi\left(\theta\right)=-l_2\left(\theta 
\mid \mathcal{H}_T\right)
\end{equation*}
iteratively using a suitable optimization algorithm. 
We have followed the example of \cite{jalilian2019etas, ogata2006space, zhuang2004analyzing} by using the Davidon-Fletcher-Powell algorithm. An alternative approach \cite{shah2020temporal} that has been used to fit some of the time-scaled models is described below.
\cite{duchesne2000alternative,duchesne2002semiparametric} and \cite{lawless2011statistical} have both proposed methods for time scaling that are alternatives to the ideal time scale. One alternative time scaling method proposed by \cite{duchesne1999multiple} is called the "calibration time scale". Calibration time scale is a time scale that is used to make a non-identifiable model identifiable by adjusting the scaling of the time axis. The method uses a transformation of the time axis, which is chosen to make the model identifiable, without changing the underlying hazard function. Another alternative is the "Proportional Hazard Time scale" \cite{kordonsky1997multiple}, which is based on the proportional hazards model. In this method, the hazard function is assumed to be proportional to a baseline hazard function and an explanatory variable. The time scale is chosen such that the coefficient of the explanatory variable is equal to one, which simplifies the interpretation of the model. \par
Here are a few examples of different time scale options \cite{kordonsky1993choice, kordonsky1995system, oakes1995multiple, sumereder2005estimation} and their corresponding assumptions:
\begin{itemize}
    \item \textbf{Ideal Time Scale:} The hazard function is constant over time. This assumption is often made when there is no clear understanding of the underlying process generating the data, or when the data is too sparse to accurately estimate a more complex hazard function. This can be represented as $t^Z=\phi_1\left(t\right)=t$, say.
    \item \textbf{Calibration Time Scale:} The hazard function is non-constant over time, but a transformation of the time axis is used to make the model identifiable without changing the underlying hazard function. The method assumes that the model is non-identifiable without the calibration time scale, but identifiable with it. Here, $t^Z=\phi_2\left(t,\ \omega\right)=\frac{t}{\omega}$, say.
    \item \textbf{Proportional Hazards Time Scale:} The hazard function is proportional to a baseline hazard function and an explanatory variable. The time scale is chosen such that the coefficient of the explanatory variable is equal to one, which simplifies the interpretation of the model. This method assumes that the hazard function is proportional to a baseline hazard function and an explanatory variable. Mathematically, $t^Z=\phi_3\left(t, Z\left(t\right)\right)=\frac{t}{Z\left(t\right)}$, say.
    \item \textbf{Log-Linear Time Scale:} The hazard function is a log-linear function of time. This method assumes that the hazard function is a log-linear function of time, which can be useful when the data exhibit a clear trend over time. Here, $t^Z=\phi_4\left(t\right)=\log{t}$, say.
    \item \textbf{Power Time Scale:} The hazard function is a power function of time. This method assumes that the hazard function is a power function of time, which can be useful when the data exhibit a clear trend over time. Hence, we have $t^Z=\phi_5\left(t, \omega\right)=t^\omega$, say.
\end{itemize}
It's important to note that the choice of time scale can have a significant impact on the results and interpretation of the model and it's essential to validate the assumptions and check the results with other models or different time scales.\par
The time-scaled ETAS model is a statistical framework that characterises the temporal and spatial distributions of aftershocks following a main shock. It modifies the time scale to facilitate the differentiation between background events and triggered events based on the intervals between occurrences. The model utilises the Epidemic Type Aftershock Sequence (ETAS) framework and incorporates the time-dependent characteristics of aftershocks. A set of factors, such as the main shock's intensity, the background rate of earthquakes, and the pace at which aftershocks decrease over time, constitute the model. The parameters are estimated from the data utilising a maximum likelihood estimation technique. A collection of spatial and temporal kernels, estimated via maximum likelihood estimation, define the spatio-temporal component of the time-scaled ETAS model, which also takes into consideration the temporal clustering of aftershocks and the spatial distribution of earthquakes.\par
We have obtained results for model performance of the ETAS model via two approaches. The first approach leads to assuming the ETAS model as a Marked Point Process and maximizing the resulting log-likelihood (read: minimizing negative log-likelihood) iteratively using the optimization procedure proposed by \cite{nelder1965simplex}. By this approach, we have, for each type of time-scaled data, considered the ETAS model as a Ground Intensity Function for two models where the underlying probability distribution of event magnitudes is either exponential or gamma, and estimates of the parameters have been thus compared. The second approach entails a similar path, except that for likelihood maximization (or negative log likelihood minimization), we opt for the Davidon – Fletcher – Powell (DFP) algorithm for parameter optimization, similar to the model proposed by \cite{ogata1988statistical, zhuang2004analyzing, ogata2006space}, coined Iterative Stochastic De-clustering Method (ISDM) in the paper \cite{jalilian2019etas}.

\section{Results}
Table \ref{tab:table6_1} compares the outputs of the ETAS model (un-scaled time) as fitted using the Iterative Stochastic De-clustering Method (assuming that event magnitudes follow a radially symmetric joint density function) and the alternative models where the ETAS model is used as a Ground Intensity Function, assuming exponential and gamma distributions of event magnitudes.\par
\begin{table}[htbp]
  \centering
  \caption{Parameter Estimates using ETAS model as Ground Intensity Function $\left\{{t}^{Z} ={\phi}_{1}\left({t}\right)={t}\right\}$}
    \begin{tabular}{|c|c|c|}
    \hline
    \textbf{Parameter} & \textbf{Exponential Model} & \textbf{Gamma Model} \\
    \hline
         $\mu$ & 0.001451 & 0.001398 \\
    \hline
         $A$ & $5.3\times 10^{-5}$ & $6.91\times 10^{-5}$ \\
    \hline
         $\alpha$ & 1.713889 & 1.647376 \\
    \hline
         $c$ & 0.540771 & 0.59556 \\
    \hline
         $p$ & 1.133662 & 1.126354 \\
    \hline
         $D$ & 0.228004 & 0.265632 \\
    \hline
         $\beta$ & {---} & 0.367643 \\
    \hline
    \textbf{Log-likelihood} & -7837.83 & -7795.65 \\
    \hline
    \end{tabular}%
  \label{tab:table6_1}%
\end{table}%
Clearly, we can see that both the models perform equally in this regard, although the Gamma model fares marginally better than the Exponential model in this case. For both the models that we see above,$\mu$ or the background seismicity rate of the process represents the average number of events occurring per unit time in the absence of any triggering effect, which is very low, indicating that more often, the earthquakes occurring in this region have been due to some triggering event. $\alpha$, the triggering rate parameter is seen to be almost identical, and is slightly higher in the exponential model than the gamma model indicating that using the gamma model may result in some relatively negligible underestimation of the triggering rate. The gamma model does have a more pronounced clustering parameter\ c, which represents the probability of an event being able to trigger one or more consequent events. The parameter\ p, or the probability distribution parameter of the model which represents the distribution of the time between an event and its triggered events, is also slightly higher in the exponential model than the gamma model. This may be attributed to the parameter $\beta$, which is responsible for the variation in the magnitude of earthquakes in the region. $D$ denotes the spatial decay parameter of the model and it represents the rate at which the triggering effect decreases with distance. As can be inferred from the table itself, we can see that they are almost identical in case of either exponential or gamma distribution of magnitudes. A may be interpreted as the minimum number of aftershocks expected in case of an earthquake at minimum threshold $m=m_0$. Again, as the estimate of\ A is very close to zero, we can expect that aftershocks are rare, and for number of aftershocks to be high, the magnitude of the triggering earthquake needs to be very high. \par
However, comparing the same model with the ETAS model fitted using ISDM and likelihood optimized using the DFP algorithm, we have the results in Table \ref{tab:table6_2}.
\begin{table}[htbp]
  \centering
  \caption{Parameter Estimates using ETAS model using ISDM optimized using DFP $\left\{{t}^{Z}={\phi}_{1}\left({t}\right)={t}\right\}$}
    \begin{tabular}{|c|c|c|}
    \hline
    \textbf{Parameter} & \textbf{Exponential Model} & \textbf{Gamma Model} \\
    \hline
         $\mu$ & 1.1153 & 0.0383 \\
    \hline
         $A$ & 0.2102 & {0.0906} \\
    \hline
         $\alpha$ & 1.5979 & {0.0263} \\
    \hline
          $c$ & 0.0071 & {0.1552} \\
    \hline
          $p$ & 1.1395 & {0.0139} \\
    \hline
          $\beta$ & 3.5912 & {0.124} \\
    \hline
          $D$ & 0.0033 & {0.191} \\
    \hline
          $q$ & 1.8398 & {0.0457} \\
    \hline
          $\gamma$ & 1.2236 & {0.073} \\
    \hline
    \textbf{Log-likelihood} & -497.874 & (AIC = 1011.749)  \\
    \hline
  \end{tabular}%
  \label{tab:table6_2}%
\end{table}%
From Table \ref{tab:table6_2}, we can safely say that the ISDM approach yielded much better results from the point of view of fit. From the interpretation viewpoint as well, we can see that this set of parameter estimates indicate significantly higher background intensity, higher value of number of aftershocks at minimum magnitude threshold (approximately 1 in 5, in this case), almost identical $\alpha$ indicating that the triggering rate estimate may be accurate. We see a significantly lower value of $c$ in this case, which may indicate low clustering rate, but we have applied an iteratively de-clustered algorithm here, so the lower value of $c$ may be justified. The $p$ parameter may be identical for both the models, indicating that the distribution of time between earthquakes may be similar, although the same cannot be said for $D$, which is lower than that of the previous model, indicating that the rate of decrease in triggering effect per unit distance is lower. Additionally, a significantly higher value of $\beta$ indicates higher variation in the magnitude over the entire area.$\gamma$ and $q$ are the shape and the scale parameters of the radially symmetric PDF $f_{D,\gamma,q}\left(x-x_i;y-y_i;m_i\right)$ respectively, and may be treated as hyperparameters, to be used for tuning the model for better accuracy. Tables \ref{tab:table6_3}, \ref{tab:table6_4}, \ref{tab:table6_5}, \ref{tab:table6_6} and \ref{tab:table6_7} are models fitted using the ETAS Ground Intensity Function with Exponential and Gamma event magnitude approach after using various time-scales discussed earlier.
\begin{table}[htbp]
  \centering
  \caption{Parameter Estimates using ETAS model as Ground Intensity Function $\left\{{t}^{Z}={\phi}_{2}\left({t},\ {\omega}\right)=\frac{{t}}{{\omega}},\ {\omega}={5}\right\}$
}
    \begin{tabular}{|c|c|c|}
    \hline
    Parameter & \textbf{Exponential Model} & \textbf{Gamma Model} \\
    \hline
    $\mu$ & 0.01451 & 0.01397782 \\
    \hline
    $A$ & 0.00053 & 0.000691442 \\
    \hline
    $\alpha$ & 1.71387 & 1.647391 \\
    \hline
    $c$ & 0.05408 & 0.05955264 \\
    \hline
    $p$ & 1.13366 & 1.126352 \\
    \hline
    $D$ & 0.228 & 0.2656316 \\
    \hline
    $\beta$ & {---} & 0.116257 \\
    \hline
    \textbf{Log-likelihood} & -5433.9 & -5391.755 \\
    \hline
    \end{tabular}%
  \label{tab:table6_3}%
\end{table}%

\begin{table}[htbp]
  \centering
  \caption{Parameter Estimates using ETAS model as Ground Intensity Function $\left\{{t}^{Z}={\phi}_{2}\left({t},\ {\omega}\right)=\frac{{t}}{{\omega}},\ {\omega}={1000}\right\}$
}
    \begin{tabular}{|c|c|c|}
    \hline
    Parameter & \textbf{Exponential Model} & \textbf{Gamma Model} \\
    \hline
    $\mu$ &  1.450546 & 1.397766  \\
    \hline
    $A$ &  0.05301076 & 0.06915157 \\
    \hline
    $\alpha$ & 1.713874 & 1.65 \\
    \hline
    $c$ & 0.000540705 & 0.000595452 \\
    \hline
    $p$ & 1.133651 & 1.126342 \\
    \hline
    $D$ & 0.2280044 & 0.2656312 \\
    \hline
    $\beta$ & {---} & 0.01162555 \\
    \hline
    \textbf{Log-likelihood} & -626.1306 & -583.9576 \\
    \hline
    \end{tabular}%
  \label{tab:table6_4}%
\end{table}%
Here, we can see that increasing the time scale to near 1000 provides better log-likelihood. However, higher values of $\omega$ have not yielded better results. One may use a search algorithm to find a scale parameter to maximize the log-likelihood, but such algorithms may require costlier processing hardware. The parameter estimates in Table \ref{tab:table6_4} is closer in interpretation to the ISDM model than the unscaled ETAS model with exponentially or gamma distributed magnitudes and also induces better fit in that regard. For the results in Table \ref{tab:table6_5}, we have used the average depth of all minor earthquakes, between two major quakes $(\text{magnitude} > 5)$, as a usage measure. \par
\begin{table}[htbp]
  \centering
  \caption{Parameter Estimates using ETAS model as Ground Intensity Function $\left\{{t}^{Z}={\phi}_{3}\left({t},\ {Z}\left({t}\right)\right)=\frac{{t}}{{Z}\left({t}\right)}\right\}$
}
    \begin{tabular}{|c|c|c|}
    \hline
    \textbf{Parameter} & \textbf{Exponential Model} & \textbf{Gamma Model} \\
    \hline
         $\mu$ & 0.0002956871 & 0.0002937223 \\
    \hline
         $A$ & 0.2932234 & 0.3480650 \\
    \hline
         $\alpha$ & 0.5707210 & 0.4834184 \\
    \hline
         $c$ & 0.03981759 & 0.05134718 \\
    \hline
         $p$ & 1.113115 & 1.117977 \\
    \hline
         $D$ & 0.2280044 & 0.3000280 \\
    \hline
         $\beta$ & {---} & 0.3114590 \\
    \hline
    \textbf{Log-likelihood} & -5148.079 & -5075.610 \\
    \hline
    \end{tabular}%
  \label{tab:table6_5}%
\end{table}%
Just like the rest of the cases, the Gamma model fares only slightly better than the Exponential model. However, the choice of the Depth variable as a usage measure is not found to be suitable for scaling purposes. The parameter estimates are closer to the unscaled ETAS model than the iteratively de-clustered ETAS model. Hence using depth as a covariate for time scaling seems unsuitable. One may opt for other usage measures like increase in height of a particular Himalayan range per unit time, which may be an interesting variable to use in this regard. Another variable that may be potentially used as a usage measure can be the rate of depletion of ice caps in the Himalayas with time. Some interesting outtakes may potentially arise from such analyses. However, collection of such data may be arduous at the ground level, as government records may not even be available in certain cases.\par
\begin{table}[htbp]
  \centering
  \caption{Parameter Estimates using ETAS model as Ground Intensity Function $\left\{{t}^{Z} ={\phi}_{1}\left({t}\right)={t}\right\}$}
    \begin{tabular}{|c|c|c|}
    \hline
    \textbf{Parameter} & \textbf{Exponential Model} & \textbf{Gamma Model} \\
    \hline
    $\mu$& 0.66566 & 0.21001 \\
    \hline
    $A$ & 0.12828 & 2.19703 \\
    \hline
    $\alpha$ & 1.36808 & 0.9307 \\
    \hline
    $c$ & 5.47117 & 0.58414 \\
    \hline
    $p$ & 0.91649 & 1.32135 \\
    \hline
    $D$ & 0.228 & 1.57623 \\
    \hline
    $\beta$ & {---} & 1.57623 \\
    \hline
    \textbf{Log-likelihood} & -4461.1 & -3637.2 \\         \hline
    \end{tabular}%
  \label{tab:table6_6}%
\end{table}%
Although the log-scale does not induce a good enough log-likelihood to compete with the calibration or proportional hazards time scale, it does fare significantly better than the unscaled model. This goes to show that a combination of other scales with the log scale may potentially provide a better model.
\begin{table}[htbp]
  \centering
  \caption{Parameter Estimates using ETAS model as Ground Intensity Function $\left\{{t}^{Z} ={\phi}_{1}\left({t}\right)={t}\right\}$}
    \begin{tabular}{|c|c|c|}
    \hline
    \textbf{Parameter} & \textbf{Exponential Model} & \textbf{Gamma Model} \\
    \hline
         $\mu$ & 0.2996577 & 0.2603446 \\
    \hline
         $A$ & 0.2401597 & 0.3450314 \\
    \hline
         $\alpha$ & 1.48376 & 1.39805 \\
    \hline
         $c$ & 0.0002384452 & 0.0002525749 \\
    \hline
         $p$ & 0.9817204 & 0.9741814 \\
    \hline
         $D$ & 0.2280044 & 0.2703550 \\
    \hline
         $\beta$ & {---} & 0.01578156 \\
    \hline
    \textbf{Log-likelihood} & -974.8525 & -928.1033 \\
    \hline
    \end{tabular}%
  \label{tab:table6_7}%
\end{table}%
The Power Scale taken at $\omega=\frac{1}{2}$ give the best result at a log-likelihood in the neighbourhood of negative one thousand, which is still better than the log-linear scale or even the proportional hazards scale. Here again, the estimates of model parameters indicate a situation closer that estimated by the iteratively de-clustered model. One may note that usage of calibration time scale in this regard, translates to easier likelihood optimization as opposed to other scales. One may use, as in our case, iterative algorithms in search of calibration parameter values that maximize log-likelihood. Using the power-scale implies that a stronger algorithm may be required to find that optimum.\par
However, the calibration time scale performs best when compared with the other time scales in terms of model fit. Having said that, the best performance has been that of the model generated using ISDM optimized using FDP algorithm. The figures provided below may provide some insight onto the earthquake patterns observed in Nepal since January 11, 1990.\par
\begin{figure}[ht]
    \centering
    \includegraphics[width=0.9\textwidth]{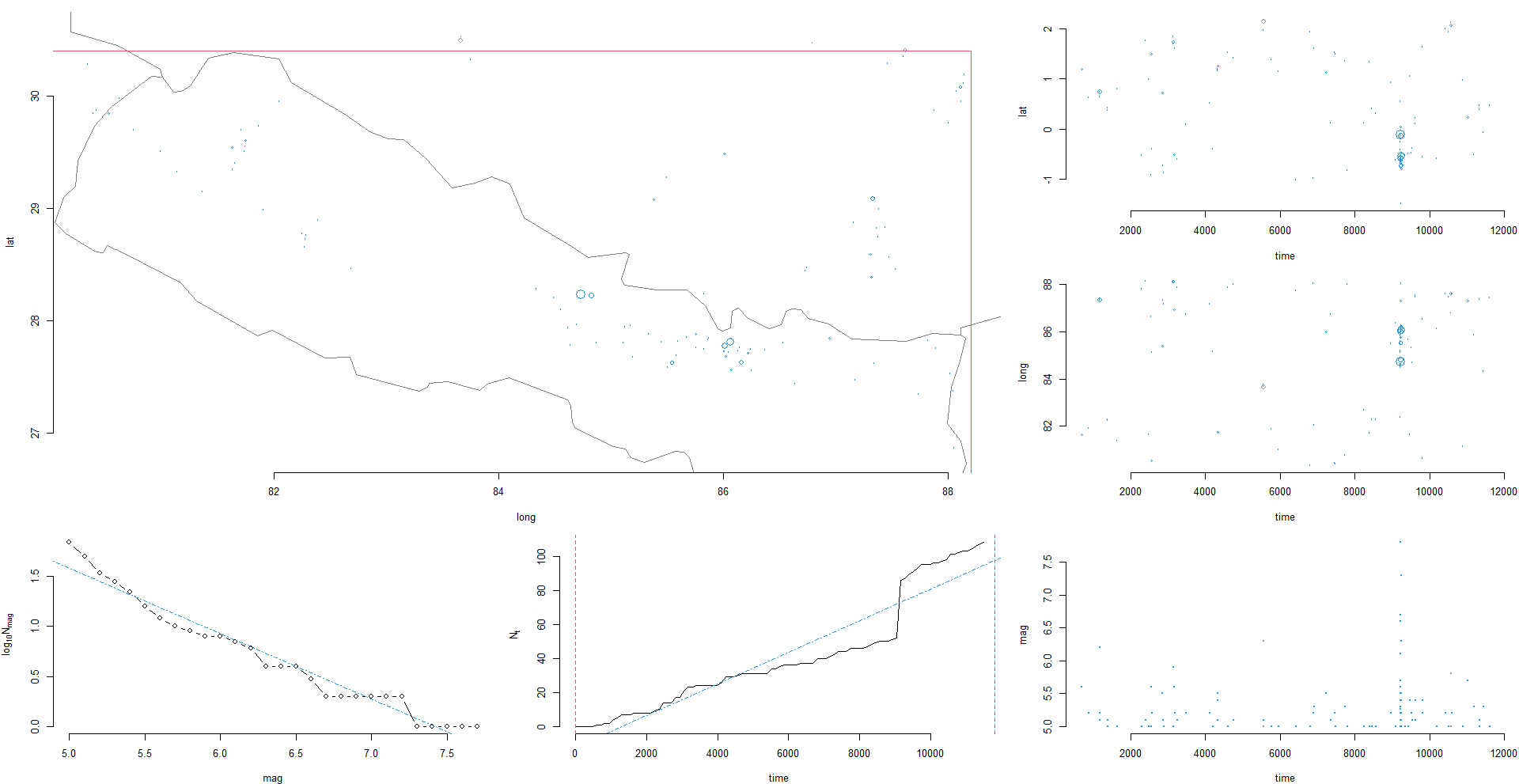}
    \caption{Location of epicentres (top-left panel), logarithm of frequency by magnitude (bottom-left panel), cumulative frequency v/s time (bottom-middle panel) and latitude, longitude and magnitude v/s time (right panels) of earthquakes with magnitude $\geq5$ occurring between 1990-01-11 and 2022-05-20 in Nepal and its proximity (27– 30 $^{\circ}N$ and 79–89 $^{\circ}\text{E}$), extracted from the ANSS global comprehensive catalogue}
    \label{fig:figure6_1}
\end{figure}
\begin{figure}[ht]
    \centering
    \includegraphics[width=0.9\textwidth]{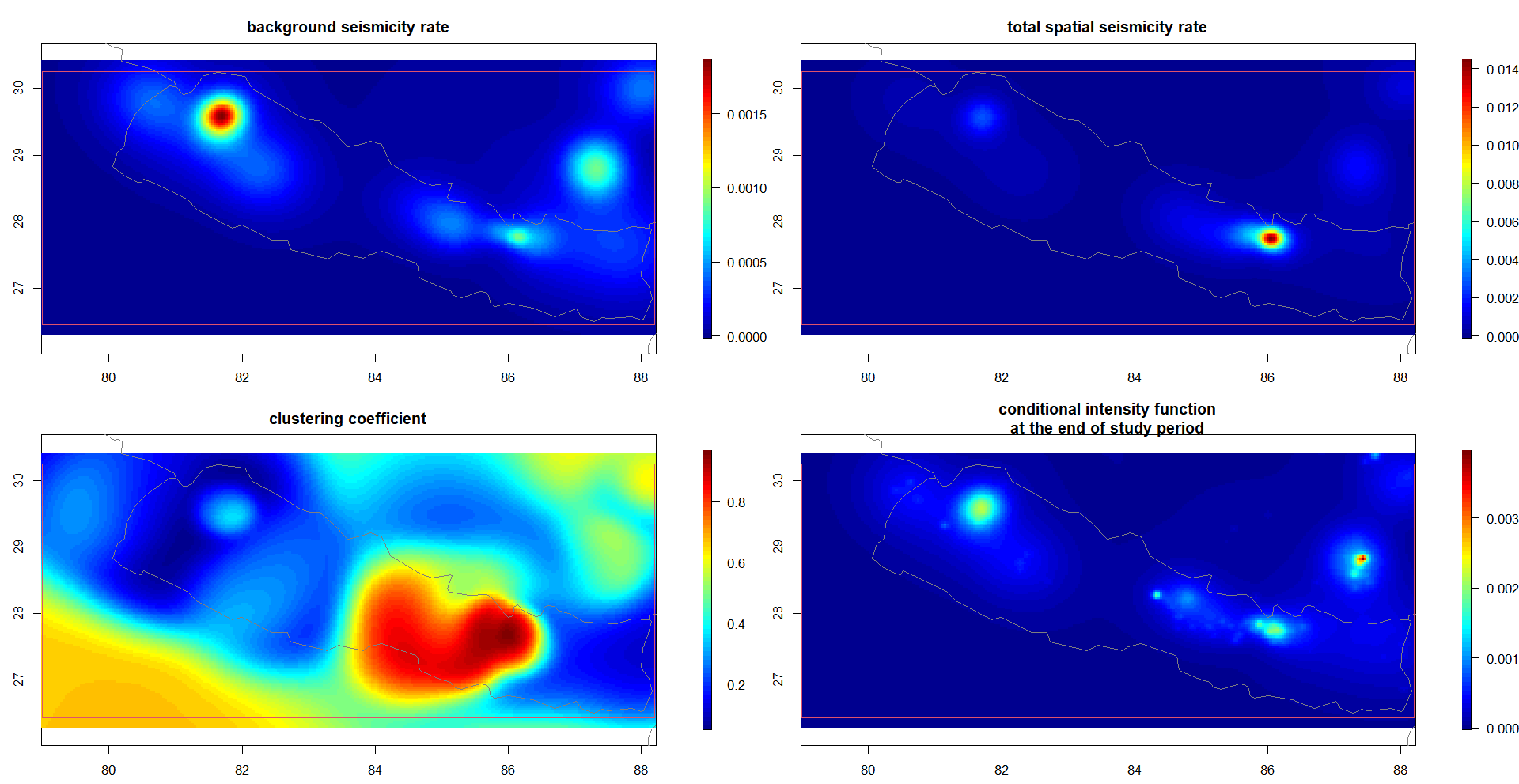}
    \caption{Plots of estimates of the background seismicity rate, total spatial intensity, clustering (triggering) coefficient and conditional intensity at $t = t_{start} + T$ for the fitted ETAS model to the Nepal catalogue}
    \label{fig:figure6_2}
\end{figure}

Fig. \ref{fig:figure6_1} indicates an outlier at  April 25, 2015, where multiple great earthquakes occurred in quick succession in the regions as indicated by the circles in the above map. 
Fig. \ref{fig:figure6_2} marks the spatial neighbourhoods where background intensity is higher and neighbourhoods that are clustering prone. 
From Fig. \ref{fig:figure6_2}, it can be said that background seismicity rate is higher in the north-western part of Nepal (Karnali region), whereas the total spatial seismicity rate and the clustering coefficient is significantly higher in central Nepal (Kathmandu and Narayani region). This indicates a higher potential of powerful aftershocks in the central Nepal region. However, earthquakes are more frequent in the North-Western region. 
\begin{figure}[ht]
    \centering
    \includegraphics[width=0.9\textwidth]{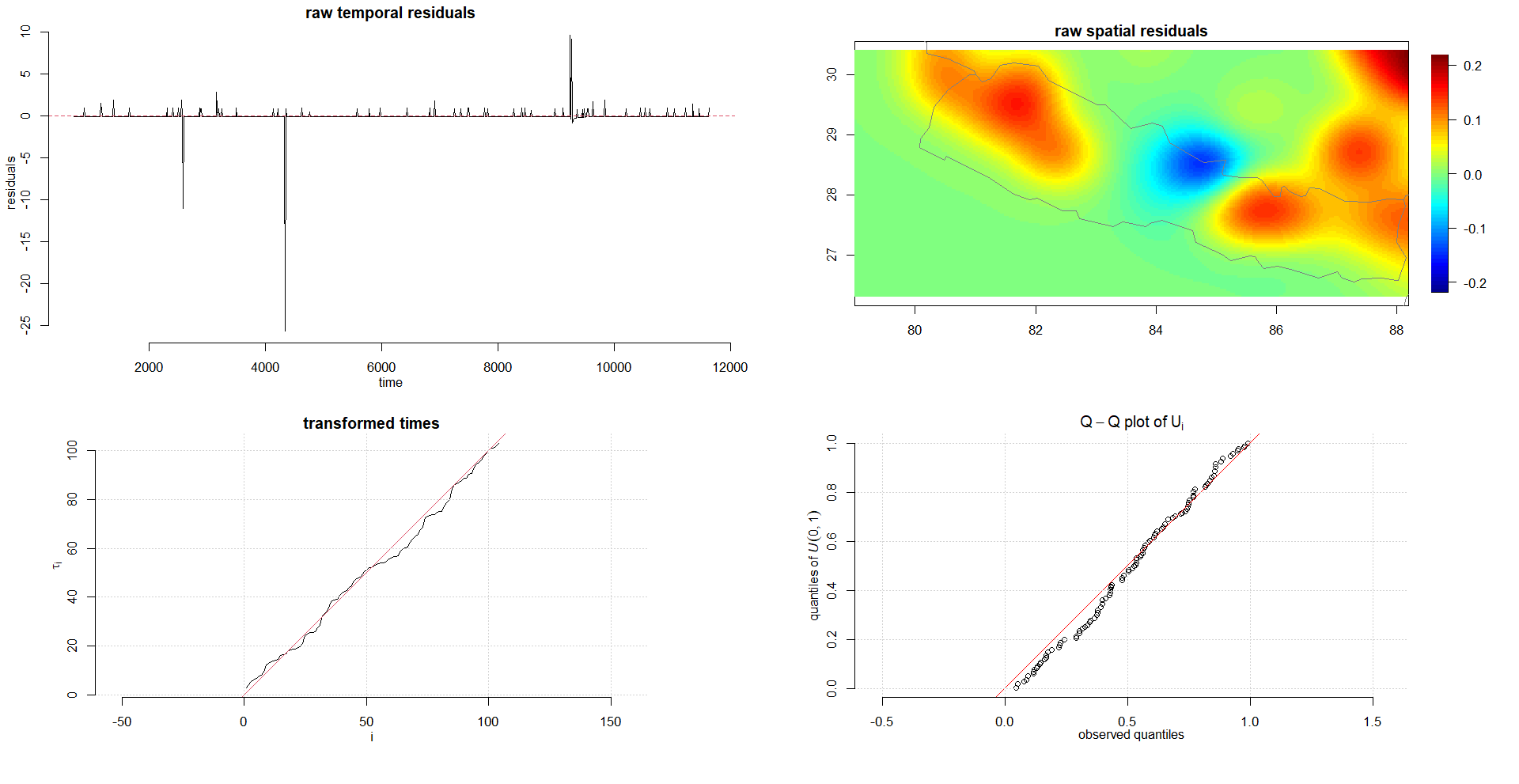}
    \caption{Diagnostic plots for the fitted ETAS model to the Nepal catalogue: the temporal residuals (top-left), smoothed spatial residuals (top-right), transformed times $\tau_i$ against $i$ (bottom-left) and the Q-Q plot of $U_i$ (bottom-right)}
    \label{fig:figure6_3}
\end{figure}
Fig. \ref{fig:figure6_3}(top-left and top-right) shows the results of model diagnostics performed on the ETAS model. It has been observed that earthquakes closer to the Kathmandu region have low residuals, meaning that the background intensity and the clustering coefficient of that region is well represented by the model. However, as we move eastwards or westwards, we see a sharp increase in the residuals. This may be the result of varying levels of terrain elevation that have not been accounted for in our models. There may also be other underlying factors that need to be investigated in greater detail. In Fig. \ref{fig:figure6_3}(bottom-left, and bottom-right), it can be seen that points on the plots of transformed times $\tau_j$ and the Q-Q plot of $U_j$ lie approximately on the $y = x$ line. To confirm, we performed a Kolmogorov-Smirnov test for goodness-of-fit. The $p$-value of $0.4386 (> 0.05)$ indicates that $U_j$ follows a $U\left(0, 1\right)$ distribution, at a $5\%$ level of significance.\par
\begin{figure}[ht]
    \centering
    \includegraphics[width=0.9\textwidth]{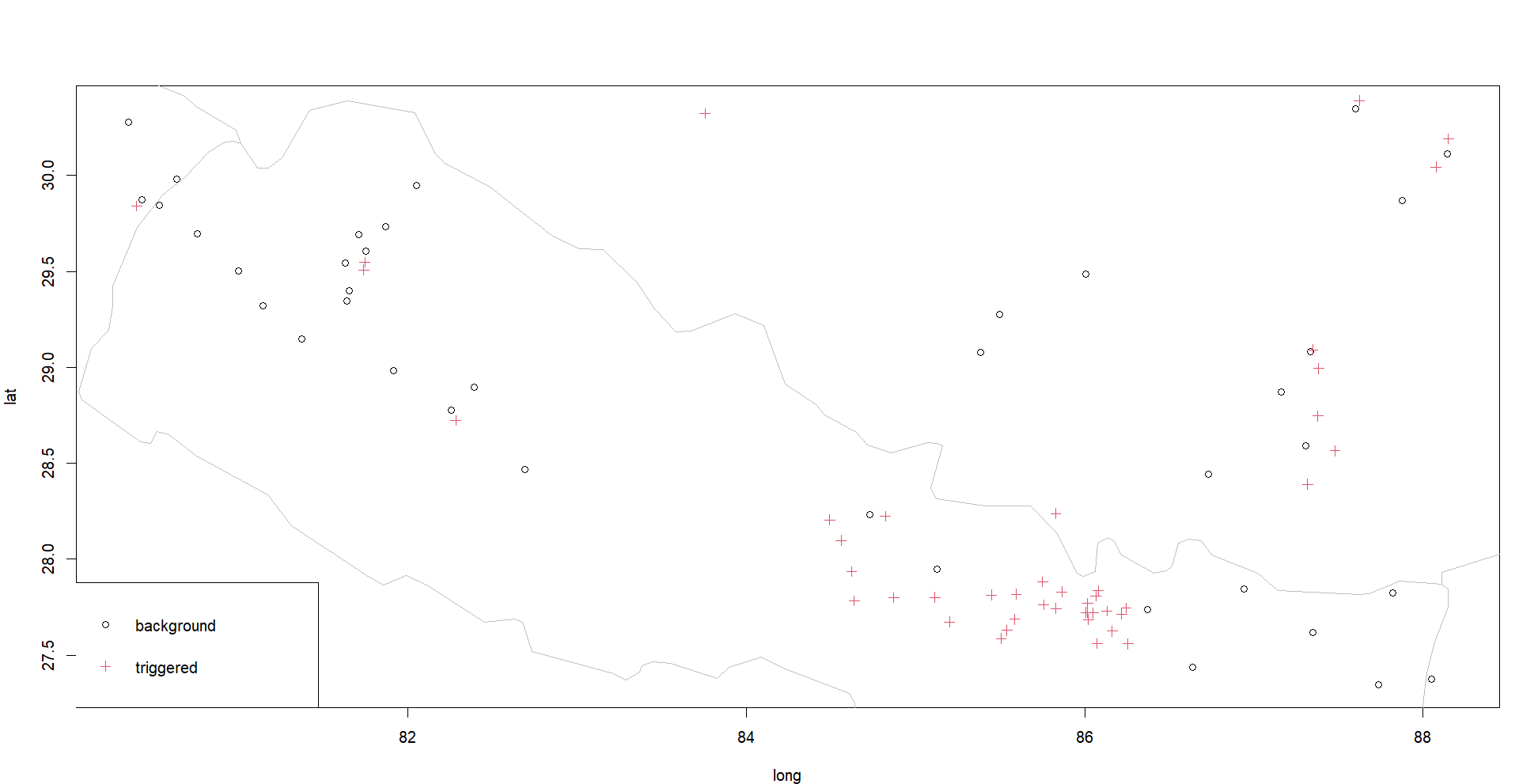}
    \caption{Most likely (Probability $> 0.95$) background and triggered events detected by estimated de-clustering probabilities $p_j$}
    \label{fig:figure6_4}
\end{figure}
Fig. \ref{fig:figure6_4} indicates that most of the quakes near the Kathmandu – Janakpur region are triggered whereas quakes occurring in the North-West region (Karnali region) mostly consist of background seismic activity.
\section{Conclusions}
From the results obtained above, the following outtakes can be derived:
\begin{itemize}
    \item The Davidon-Powell-Fletcher algorithm, fares better than the Nelder-Mead Optimization algorithm for maximum likelihood estimation in this case. Additionally, the probability distribution of the magnitude, if taken to be either exponential or gamma, fits very poorly on the data. The radially symmetric density assumption on the other hand yields a superior fit.
    \item Judging by interpretability, we should take into account, the fact that, in case of the exponential or gamma assumption of magnitude distribution, the inherent interpretability of the model is increased, which loosely translates to better foundations for decision-making. The parameters of the radially symmetric density are complex to interpret and to base decisions on. However, they may be used as hyper-parameters in “\emph{deep}”-er models, for obtaining better forecasts.
    \item Although by a very small margin, among the theoretical distributions that have been postulated for magnitude probability distribution, the gamma model fares better in all types of time scales than the exponential model. That may be due to an extra parameter which makes the difference. However, having said that, the cost of including an extra parameter may prove costlier in terms of likelihood per additional covariate, if too many instances are to be trained to the model.
    \item The residual map indicates lower values of residuals in the central part of the country whereas as we go from the centre outwards, residuals become higher. This may be a direct connection with the extent of urbanization in those areas to the magnitude and frequency. The areas having low variance in raw spatial residuals, are the more urbanized areas of the country, whereas relatively less inhabited areas have very high variation of magnitude. This may be solved by scaling the magnitudes along with time.
    \item The conditional density and the clustering coefficient for central locations of the country have a higher probability of aftershocks than the locations, and as we go eastwards or westwards from the centre. One may take the usage measure as some variable highly correlated with the magnitude for this purpose.
\end{itemize}
Although a lot of research in the sphere of seismological modelling is centered around exploiting the flexibility offered by the ETAS model to accommodate additional constraints or to bypass assumptions so as to introduce robustness, as evident by the recent works \cite{altun2021new, davoudi2020aftershock, hirose2021characteristics, hong2021multi, motagi2023point, tavakoli2018aftershock}, and the recent advances in the domain of time-scaling for lifetime data \cite{lawless2010models, nekrasova2023seismic, syamsundar2020alternative}, collaboration between researchers in these two spheres seldom occurs, resulting in dearth of opportunities and unexplored possibilities. Ours seems to be one of the earlier attempts at bridging these two domains in comprehensive manner. However, the future seems hopeful, as there lies tremendous prospect in cross-collaborations with the computational domain as well, given that research in deep modelling is at the peak, with deep models for both, seismicity as well as for time-between-event responses, is now commonplace.

\section*{Acknowledgments}
This was was supported in part by the Department of Statistics, Faculty of Sciences, The Maharaja Sayajirao University of Baroda, Vadodara, Gujarat, India, PIN--390002.

\bibliographystyle{unsrt}  
\bibliography{references}

\end{document}